\documentclass[twocolumn,prl,amsmath,letterpaper,floatfix,showpacs]{revtex4}
\usepackage{graphicx}
\usepackage{amsmath}
\usepackage{amssymb}
\usepackage{color}

\newcommand{\us}{{$\mu$s }}
\newcommand{\um}{{$\mu$m }}

\begin{document}

\title{From gyroscopic to thermal motion: a crossover in the dynamics of molecular superrotors}

\author{A. A. Milner, A. Korobenko, K. Rezaiezadeh, V. Milner}
\date{\today}

\affiliation{Department of  Physics \& Astronomy, The University of British Columbia, Vancouver, Canada}

\begin{abstract}
Localized heating of a gas by intense laser pulses leads to interesting acoustic, hydrodynamic and optical effects with numerous applications in science and technology, including controlled wave guiding and remote atmosphere sensing. Rotational excitation of molecules can serve as the energy source for raising the gas temperature. Here, we study the dynamics of energy transfer from the molecular rotation to heat. By optically imaging a cloud of molecular superrotors, created with an optical centrifuge, we experimentally identify two separate and qualitatively different stages of its evolution. The first non-equilibrium ``gyroscopic'' stage is characterized by the modified optical properties of the centrifuged gas - its refractive index and optical birefringence, owing to the ultrafast directional molecular rotation, which survives tens of collisions. The loss of rotational directionality is found to overlap with the release of rotational energy to heat, which triggers the second stage of thermal expansion. The crossover between anisotropic rotational and isotropic thermal regimes is in agreement with recent theoretical predictions and our hydrodynamic calculations.
\end{abstract}

\maketitle

Intense ultrashort laser pulses can deposit energy in a transparent gas medium either by ionizing\cite{Tzortzakis2001, Filin2009} or spinning up\cite{Steinitz2012, Zahedpour2014} its molecular constituents. Heating of the gas is followed by the formation of an acoustic wave\cite{Yu2003, Kartashov2006, Kiselev2011, Milner2015b} which leaves behind a long-lived low-density depression channel\cite{Cheng2013}. Laser-induced sound emission is key to photoacoustic spectroscopy\cite{Siebert1980, West1983, Schippers2011}, whereas laser control of the gas density proved valuable for generating\cite{Lahav2014, Jhajj2014} and even controlling\cite{Zahedpour2014} wave guiding channels in ambient air. Because ionization-free delivery of energy by means of rotational excitation does not inhibit coherent light propagation through the heated gas, it has the potential of producing longer wave guides, e.g. if executed inside ionization-free filaments\cite{Bejot2010}.

Together with the importance of rotational heating, recent developments in controlling molecular rotation with nonresonant laser pulses\cite{Ohshima2010, Fleischer2012} have stimulated active research on the exchange of energy between a rotating molecule and its environment. Transient molecular alignment has been first proposed\cite{Ramakrishna2006} and later implemented as a powerful probe of collisional relaxation in a series of pioneering experiments\cite{Houzet2012b, Vieillard2013, Karras2014}. The three relaxation steps, associated with (i) rotational decoherence, (ii) rotational reorientation and (iii) rotation-translation (RT) thermalization, have been identified in the theoretical model\cite{Hartmann2012}, yet found to overlap in time too closely to enable an individual experimental study of each process separately\cite{Karras2014}.

Extreme rotational excitation of molecular superrotors\cite{Korobenko2014a} with an optical centrifuge\cite{Karczmarek1999, Villeneuve2000} allowed us to separate the time scales of the relaxation mechanisms and investigate them in great detail. In our earlier work\cite{Milner2014a, Milner2014b}, we have focused on the rotational decoherence of superrotors which takes a few hundred picoseconds under ambient conditions and ends well before the onset of the rotational reorientation. Ultrahigh values of the molecular angular momentum $J$, provided by the centrifuge, together with the propensity of collisions to conserve its orientation \cite{Hartmann2012} results in the ``gyroscopic stage''\cite{Khodorkovsky2015}, which outlives the rotational coherence by a few nanoseconds. Here, we study the dynamics of this relaxation step by tracking an optical birefringence of the centrifuged gas in the direction perpendicular to the direction of the centrifuge. This \textit{transverse} birefringence stems from a strong permanent confinement of molecular superrotors in the plane of their rotation. In the \textit{longitudinal} direction along the centrifuge, the change in refractive index owing to the fast unidirectional molecular rotation creates a refractive ``gyroscopic channel'', also detected in this work by means of the phase contrast imaging.

As the superrotors undergo collisional relaxation, their rotational energy is released, driving the temperature of the gas up. The rotation-translation energy transfer is accompanied by a crossover from the non-equilibrium gyroscopic to the thermal phase of molecular dynamics, which to the best of our knowledge is observed here for the first time. Hydrodynamics of a locally heated gas result in the corresponding change in its pressure, followed by the emission of an acoustic wave and the formation of a low-density thermal channel. By directly measuring the evolution of the gas density on the time scale of tens and hundreds of nanoseconds, following its thermal expansion for up to 3 microseconds, we are able to project it back to the origin of the thermal phase and analyze the heating rate at the RT crossover. The analysis, based on our full hydrodynamic calculations, points at the close temporal overlap between the rotational reorientation and rotational cooling, in accord with recent theoretical predictions\cite{Khodorkovsky2015}.

\begin{figure*}
  \includegraphics[width=1.8\columnwidth]{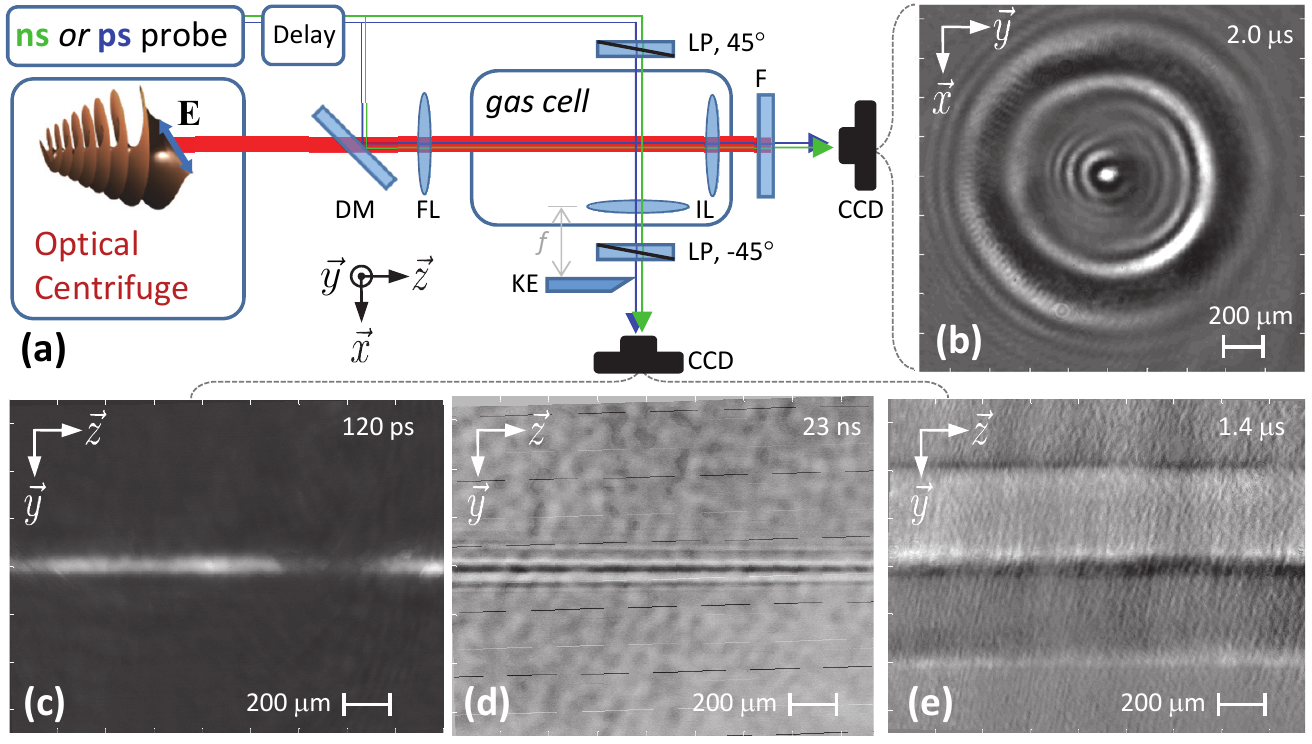}\\
  \caption{Schematic diagram of the experimental setup. An ensemble of molecular superrotors is created in a gas cell with an optical centrifuge (red beam). Nanosecond or picosecond probe pulses (green and blue beams, respectively) are delayed with respect to the centrifuge pulses and propagate either collinear with, or perpendicular to the centrifuge, creating an image of the rotationally excited volume of gas on a CCD camera either in the longitudinal or transverse direction, respectively. DM: dichroic mirror, FL: focusing lens, IL: imaging lens, KE: knife edge (shown along $y$ instead of $z$ axis for clarity), LP: linear polarizers at $\pm45^{\circ}$ to $y$ axis, F: frequency filter. An example of the longitudinal image with a circularly expanding sound wave is shown in the upper right corner. Images in the lower row were taken in the transverse geometry at early (left), intermediate (middle) and late (right) time moments. The leftmost image was recorded with the two linear crossed polarizers in place. The rightmost schlieren image was recorded with the knife edge in place.}
\label{Fig-setup}
\end{figure*}
Our experimental configuration is depicted in Fig.\ref{Fig-setup}. The details of our optical centrifuge have been discussed in an earlier publication\cite{Milner2014b}. Briefly, the centrifuge shaper, built according to the original recipe of Karczmarek \textit{et al.} \cite{Karczmarek1999}, is followed by a home built Ti:Sapphire multi-pass amplifier boosting the pulse energy to 30 mJ. Centrifuge pulses are about 100 ps long and their linear polarization undergoes an accelerated rotation, reaching the angular frequency of 10 THz by the end of the pulse. The centrifuge beam is focused by a weak 1 m focal length lens down to a spot size of 90 $\mu$m diameter (full width at half maximum) inside a gas cell filled with 0.9 bar of oxygen. This focusing results in the peak intensity of $2.2\times 10^{12}$ W/cm$^{2}$, significantly lower than the ionization threshold of O$_{2}$. We monitor the performance of the centrifuge by means of coherent Raman spectroscopy of the created molecular superrotors. The Raman setup is described elsewhere\cite{Korobenko2014a} and not shown in Fig.\ref{Fig-setup} for clarity.

We image the volume of the centrifuged gas both on \textit{short} and \textit{long time scales}, $\tau <23$ ns and $\tau <2$ $\mu$s, respectively. For a short-time scan, picosecond pulses are extracted from the same Ti:Sapphire ultrafast system as the centrifuge, and delayed with femtosecond precision by means of computer controlled motorized translation stages. Long-time scans are executed with nanosecond pulses from a separate YAG laser, while their delay is controlled electronically. In both cases, single-shot images are recorded with a CCD camera and typically averaged over 400 laser pulses.

Two imaging geometries are implemented: \textit{longitudinal} and \textit{transverse}, with probe pulses propagating collinear with, or at 90 degrees to the direction of the centrifuge. A sample longitudinal image, taken 2 $\mu $s after the centrifuge pulse, is shown in the upper right corner of Fig.\ref{Fig-setup}. The location of the central bright spot, surrounded by multiple circular interference fringes, corresponds to the position of the centrifuge beam. Because of the centimeter-long confocal parameter, the fringe pattern stems from the diffraction of probe pulses inside the long density depression channel and does not reflect the true distribution of light intensity (e.g. wave guiding) or refractive index\cite{Wahlstrand2014}. Yet owing to the linearity of the weak probe propagation and its extended length, the fringe contrast serves as a sensitive indicator of the centrifuge-induced changes in the refractive index of the gas, $\Delta n$.

\begin{figure}
  \includegraphics[width=.8\columnwidth]{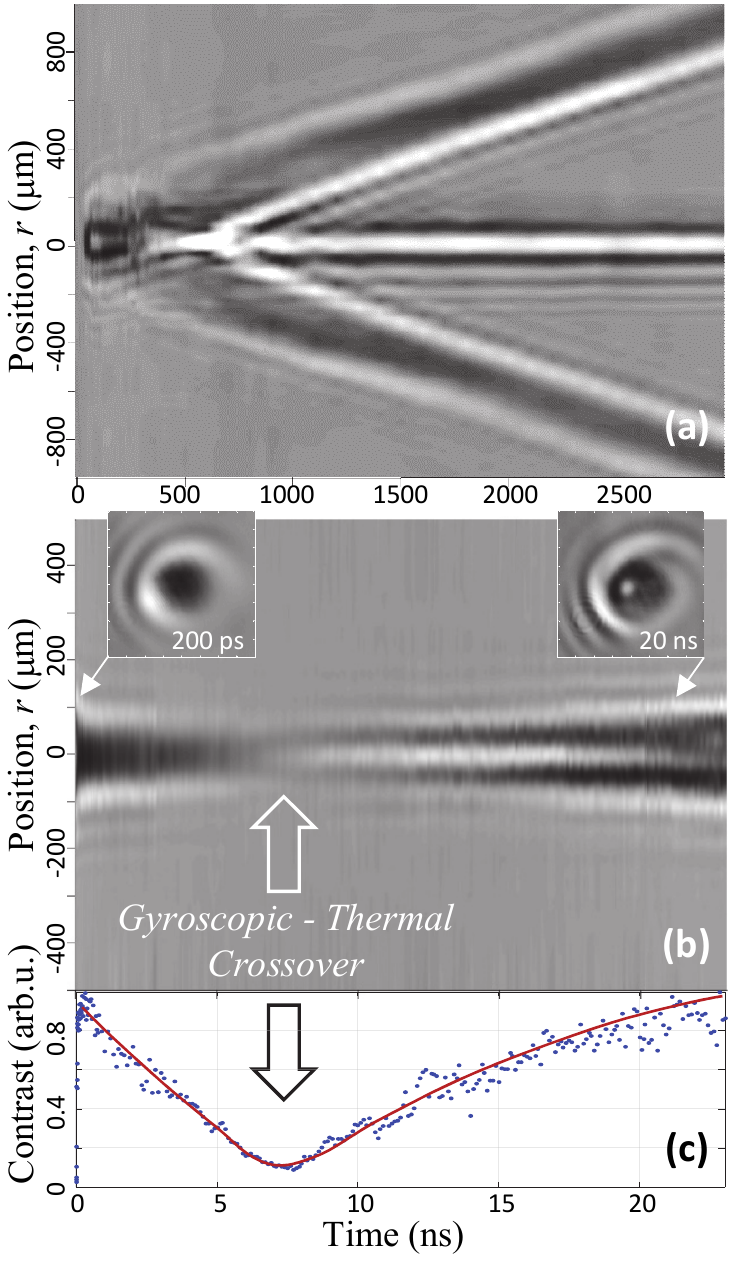}\\
  \caption{Radial cross-section of images recorded along the centrifuge beam (as in Fig.\ref{Fig-setup}(\textbf{b})) as a function of the delay time between the centrifuge and probe pulses. Long-time scan in panel (\textbf{a}), carried out with ns probe pulses, extends to 3 \us and shows an acoustic wave traveling with the speed of sound away from the central core (see Movie 1 in Supplementary Material). In panel (\textbf{b}), we zoom in on the first 23 nanoseconds (by switching to ps probe pulses) and find a crossover between the rotation-induced and thermal channels around 7 ns, when the image contrast vanishes almost entirely. Sample images of both refractive channels are shown in the square insets. The crossover is quantified by calculating the change in the image contrast and plotted in panel (\textbf{c}). The red line on top of the experimental data points is shown to guide the eye.}
\label{Fig-longitudinal}
\end{figure}
We analyze the time dependence of $\Delta n$ by recording a series of longitudinal images while scanning the delay time between the centrifuge and probe pulses. An angle-averaged radial cross-section of the recorded interference pattern is calculated and plotted as a function of the delay in Fig.\ref{Fig-longitudinal}(\textbf{a}). The plot shows a permanent central core and an outgoing sound wave, similar to the recently observed dynamics of plasma filaments\cite{Wahlstrand2014, Lahav2014}. Unlike the case of plasma, however, zooming in on the first few nanoseconds of the filament formation reveals unique details of its early history. The instantaneous creation of a refractive channel by the centrifuge, as seen in panel (\textbf{b}) of Fig.\ref{Fig-longitudinal}, reflects its non-thermal origin. Indeed, we detect the signal already in the first picoseconds after the arrival of a centrifuge pulse - too short a time scale to allow collisional thermalization. Instead, the collisions seem to suppress, rather than enhance $\Delta n$, which disappears almost completely at around 7 ns. From that moment on, the channel grows back and eventually emits the sound wave seen in Fig.\ref{Fig-longitudinal}(\textbf{a}), in accord with the expected thermal dynamics. A clear crossover between the two regions is illustrated in panel (\textbf{c}) of Fig.\ref{Fig-longitudinal}, where we plot the contrast of the recorded longitudinal images as a function of time. To identify the nature of the refractive channels before and after the crossover, we switch to the transverse imaging geometry.

Three typical transverse images, recorded on three qualitatively different time scales, are presented at the bottom of Fig.\ref{Fig-setup}. The schlieren technique\cite{SchlierenBook} is employed by inserting a knife edge in the focal plane of the imaging lens, to determine the true distribution of the refractive index (more exactly, its d/d$y$ derivative) at long delay times. The side view of a narrow channel is free of interference fringes, seen in the longitudinal geometry, which greatly simplifies the interpretation of the picture. Schlieren image (\textbf{e}) on the right shows two high-density waves captured 400 \um above and below the low-density depression channel in the center. The snapshot (\textbf{d}) in the middle was taken close to the very origin of the thermal channel around $\tau\approx 20$ ns, before the beginning of the hydrodynamic expansion of the gas. At even shorter delay times of order of, and below, 5 nanoseconds, our phase-contrast imaging approaches its signal-to-noise limit. At these early moments, we re-gain sensitivity by taking the images between two crossed linear polarizers, probing the centrifuge-induced birefringence of the gas (leftmost picture (\textbf{c}) at the bottom of Fig.\ref{Fig-setup}).

\begin{figure}
  \includegraphics[width=1\columnwidth]{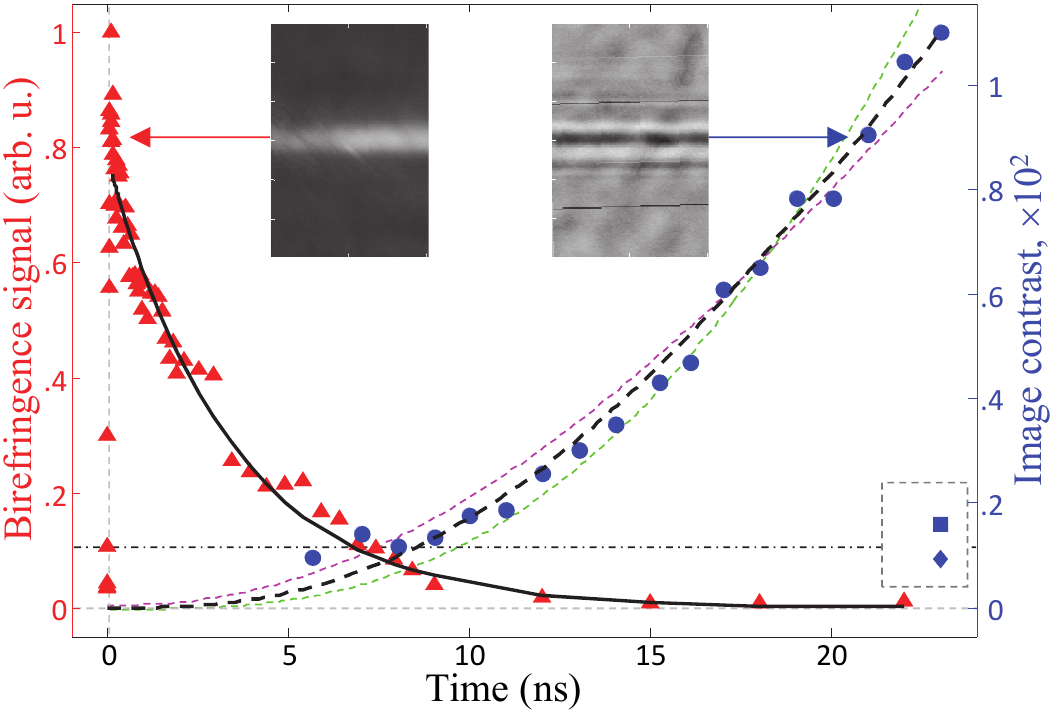}\\
  \caption{Birefringence signal (red triangles) and phase contrast (blue circles) of the centrifuged gas, retrieved from the transverse images taken with and without the crossed polarizers (left and right insets, respectively) and plotted as a function of the delay time between the centrifuge and probe pulses. Optical birefringence is induced by the centrifuge instantaneously and falls off exponentially with a time constant of 3.4 ns (solid black curve, fit at $\tau \geq 100$ ps, i.e. after the end of the centrifuge pulse). A crossover to the growing phase contrast is observed around 7 ns, similarly to the crossover in Fig.\ref{Fig-longitudinal}. Two data points in the lower right corner (within the dashed rectangle) indicate the drop in the phase contrast for the slower rotating centrifuge. The dashed black line is a fit to the hydrodynamic numerical calculations of $\Delta n(\tau )$ with an RT exchange rate of 3.4 ns, whereas the dashed magenta and green lines represent the best fits with that rate constant being five times shorter and longer, respectively. The horizontal dash-dotted line indicates the noise floor for the phase contrast measurement.}
\label{Fig-transverse}
\end{figure}

The difference between the two processes affecting optical properties of the centrifuged gas before and after the observed crossover becomes apparent when we compare the time dependence of the detected optical birefringence with that of the phase contrast. The two observables, extracted from the images taken with and without the crossed polarizers, respectively, are plotted in Fig.\ref{Fig-transverse}. The birefringence is expressed in arbitrary units with 1 being equivalent to the polarization rotation angle of about 1 mrad, while the phase contrast is quantified as the image intensity contrast. From these results, the disappearing channel to the left of the crossover in Fig.\ref{Fig-longitudinal}(\textbf{b}) can be clearly correlated with the decaying birefringence signal, and therefore associated with an optical anisotropy induced by the directional molecular rotation. As the molecular superrotors lose their rotational energy to heat, this \textit{gyroscopic channel} dies off while thermal effects take over, causing an isotropic change in the gas density reflected in the phase contrast images at $\tau \gtrsim 7$ ns. To make sure that superrotors are responsible for the observed optical changes, we repeat the experiments with the centrifuge pulses of the same peak intensity but lower terminal rotation frequency. Two points inside the dashed rectangle in Fig.\ref{Fig-transverse}, upper square and lower diamond, correspond to the centrifuge producing only slow rotors and no rotors at all, respectively.

To verify the mechanism behind the rotation-translation crossover, we numerically simulate the hydrodynamics of the ideal gas exposed to a known heat source (see Methods). The hydrodynamics equations, corresponding to the conservation of mass, momentum and energy in the compressible gas flow, are solved for the gas density $\rho$, velocity $\mathbf{v}$ and pressure $P$, assuming cylindrical symmetry and equilibrium initial and boundary conditions. Rotation-translation energy exchange is modeled by adding an external pressure source, $\frac{1}{P_{0}} \left( \partial P / \partial t\right)_\text{RT} = \frac{\Delta T}{T_{0}} \frac{1}{\tau _{b}} e^{-t/\tau_{b}}$, to the corresponding differential equation for $P$. As we argue in Methods, an exponential increase of the gas temperature (and hence, its pressure in the first few nanoseconds) with the same time constant $\tau _{b}=3.4$ ns as the decay of the birefringence signal, stems from the proportionality of both the birefringence signal and the rotational energy to $J^{2}$. In the expression above, $T_{0}$ ($P_{0}$) is the ambient temperature (pressure) and $\Delta T$ is the temperature increase at the end of the thermalization process. To estimate $\Delta T$, we note that the peak intensity of our centrifuge pulses is sufficient to spin adiabatically only 2\% of O$_{2}$ molecules occupying the lowest rotational state at room temperature. After these molecules are excited to $J\approx91$ by the centrifuge (as determined by means of Raman spectroscopy\cite{Korobenko2014a}), each oxygen superrotor carries 1.5 eV of rotational energy. Redistributing this energy among the whole molecular ensemble and all 5 degrees of freedom results in $\Delta T=140$~K.

\begin{figure*}[!t]
  \includegraphics[width=1.9\columnwidth]{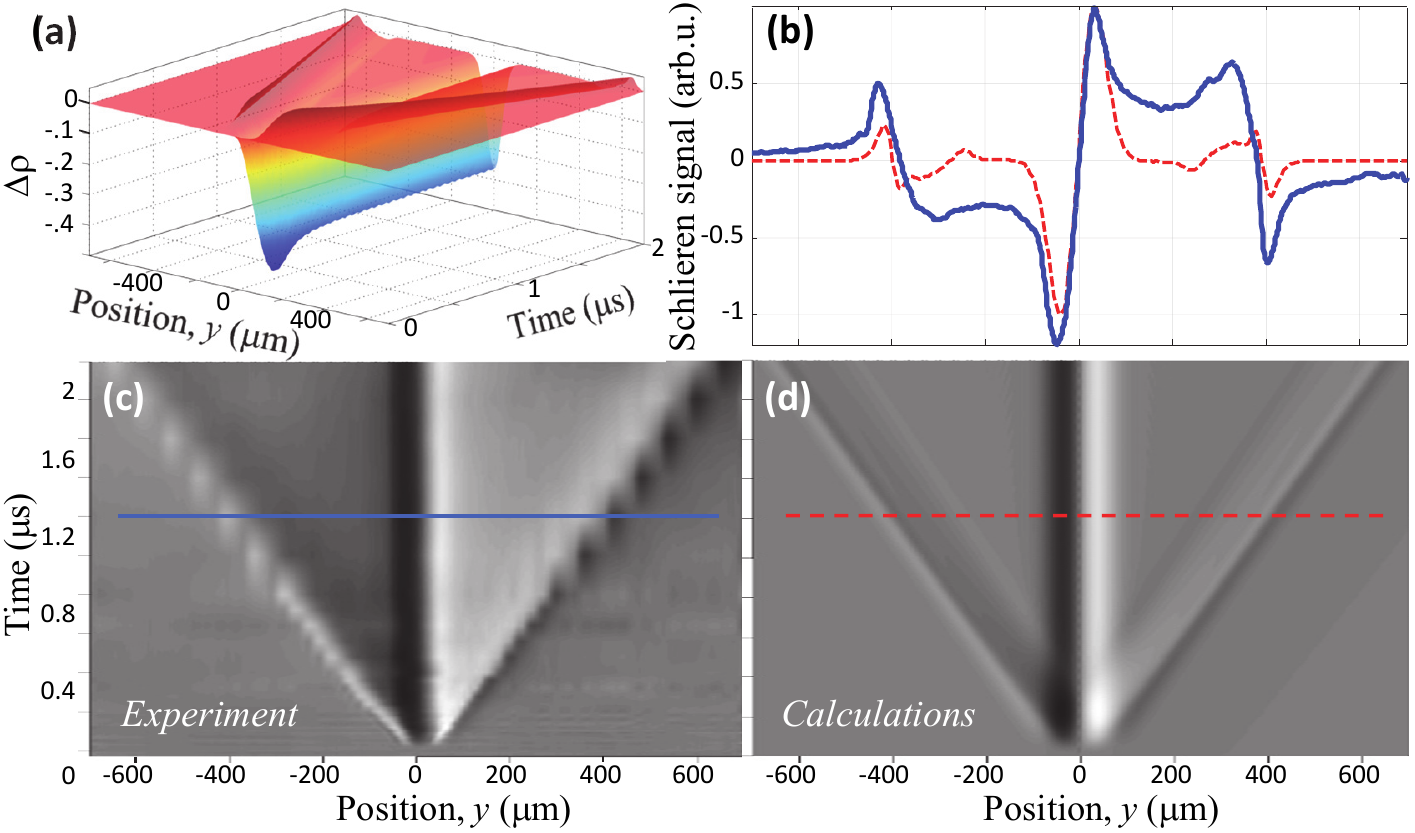}\\
  \caption{Comparison of the experimental results with the numerical calculations of gas hydrodynamics. The calculated change in the gas density, $\Delta \rho $, is shown in panel (\textbf{a}) as a function of time and distance. Experimentally determined parameters of the observed gyroscopic channel are used to simulate the heat source, which initiates the dynamics at point ($y=0, \tau =0$). In panel (\textbf{b}), the derivative of the calculated density profile d/d$y[\rho (y)]$ at 1.2 \us is compared with the $y$ cross-section of the schlieren image recorded 1.2 \us after the centrifuge (dashed red and solid blue curves, respectively). The dependence of the measured and calculated schlieren signals on both space and time is shown in panels (\textbf{c}) and (\textbf{d}), respectively, with the two lines indicating the two cross-sections displayed in plot (\textbf{b}).}
\label{Fig-analysis}
\end{figure*}
The main result of our calculations is shown in Fig.\ref{Fig-analysis}(\textbf{a}), where the change in the gas density, $\Delta \rho (y,\tau )$, is plotted as a function of both the distance from the centrifuge and the time since its arrival. One can see the density depression channel forming in the center in the first 200 ns and an acoustic density wave spreading radially away from it. Given the linear dependence of the refractive index of a gas on its density, we compare the experimentally observed phase contrast with the numerically calculated $|\Delta \rho (y=0,\tau )|$ in Fig.\ref{Fig-transverse} (blue circles and dashed black line, respectively). The only fitting parameter is the proportionality coefficient between the gas density and the image contrast. An excellent agreement between the experimental and numerical line shapes supports our interpretation of the gyroscopic channel and its role as the energy source for the thermal gas expansion. If the RT exchange rate is assumed to be very different from the birefringence decay rate $\tau _{b}$, the numerical results do not fit the experimental data as well (dashed magenta and green lines in Fig.\ref{Fig-transverse}).

Further comparison between our experimental findings and numerical model is presented in Fig.\ref{Fig-analysis}. In panel (\textbf{b}), the schlieren signal at 1.2 \us is calculated by taking the derivative of $\Delta \rho (y,\tau =1.2\text{ $\mu$s})$ with respect to $y$, and plotted together with the $y$ cross-section of the experimental schlieren image taken at that delay time (cf. Fig.\ref{Fig-setup}(\textbf{e})). The dependence of schlieren signals on both $y$ and $\tau $ is shown in Fig.\ref{Fig-analysis}(\textbf{c}) and (\textbf{d}) for the experimental and numerical   data, respectively. The experimental scan, carried out with ns probe pulses, extends to 2 \us and shows an acoustic wave traveling with the speed of sound away from the central core (see Movie 2 in Supplementary Material). The equivalent numerical scan displays very similar behavior, supporting the validity of the model.

To summarize, we report on the experimental observation of a refractive ``gyroscopic channel'', created by an ultrahigh rotational excitation of molecules with an optical centrifuge. Owing to the non-thermal nature of the channel, it affects the optical properties of a gas, such as its index of refraction and birefringence, on ultrashort (picosecond) time scale which may prove useful for ultrafast optical switching.  The gyroscopic channel is found to serve as a precursor to the formation of a conventional thermal channel (typically originating from the laser-induced plasma at much higher laser intensities), and a supplier of energy to the latter. A crossover between the rotation relaxation and thermal expansion, driven by the rotation-translation energy exchange, is discovered experimentally and reproduced in the numerical gas hydrodynamics calculations.

Providing a few electron-Volts of rotational energy per molecular superrotor, comparable to the energy of free electrons in plasma filaments, an optical centrifuge is shown to be a non-ionizing alternative for initiating and studying thermal processes in ambient gases, e.g. photo-acoustics\cite{Siebert1980} or wave guiding\cite{Lahav2014, Jhajj2014}, especially valuable because of its capability to precisely control the amount of energy transferred to the gas\cite{Korobenko2014a}. Translated to the control of the optical properties of gases, demonstrated here, this may be of use in applications ranging from ultrafast polarization switching\cite{Heritage1975, Hartinger2006, Marceau2009} to remote atmosphere sensing\cite{Zheltikov2012}. Ultrafast rotation of centrifuged molecules results in the increasing adiabaticity of molecular collisions and correspondingly slower release of rotational energy. A series of fascinating phenomena, such as an explosive thermalization and anisotropic diffusion, have recently been predicted for a gas with high concentration of molecular superrotors\cite{Khodorkovsky2015}. This work is a step towards exploring this new area of gas dynamics.

\section{Methods}
In this work, we numerically solve the following set of hydrodynamic equations\cite{LL-Hydrodynamics}:
\begin{eqnarray*}
\partial/\partial t (\rho)  & = & -\nabla \left[ \rho \mathbf{v} \right], \\
\partial/\partial t (\mathbf{p})  & = & -\nabla \left[ \mathbf{p} \mathbf{v} +P \right], \\
\partial/\partial t (\mathcal{E})  & = & -\nabla \left[ \mathcal{E} \mathbf{v} +P \mathbf{v} + \mathbf{q} \right],
\end{eqnarray*}
where $\rho$, $\mathbf{p}= \rho\mathbf{v} $ and $\mathcal{E}= \frac{1}{2}\rho v^{2}+\varepsilon$ are the mass, momentum and energy of the unit gas volume; $\mathbf{v}$ its velocity and $\varepsilon=\rho c_\text{v} T$ its internal energy, with $c_\text{v}$ being the heat capacity at constant volume and T the gas temperature. $P$ is the pressure of the gas and $\mathbf{q}= -\kappa \nabla T$ is the heat flux, with $\kappa $ being the heat conductivity.

An optical birefringence owing to the planar confinement of diatomic molecules has been recently analyzed and described in terms of the expectation value of $\cos ^{2}\theta _{z}$, where $\theta _{z}$ is the angle between the axis of a molecule and the propagation axis $z$ of probe light\cite{Hoque2011}. The birefringence signal is proportional to $\left( 1 - 3\langle \cos^{2}\theta_{z} \rangle \right)^2$, becoming non-zero when the molecules are confined to the $xy$ plane ($\langle \cos^{2}\theta_{z} \rangle < 1/3$).  For a gas of optically centrifuged molecules in the quantum state $|J, M_{J}=J \rangle$, where $M_{J}$ is the projection of $\mathbf{J}$ on $\hat{z}$, one finds\cite{KremsBook}: $\langle \cos^{2}\theta_{z} \rangle = \langle J,J| \cos^{2}\theta_{z} |J,J \rangle = (2J+3)^{-1}$. As a result, the birefringence signal is proportional to $J^{2}$. Since the rotational energy scales with the same power of $J$, one can show that the growth of the gas temperature due to the rotation-translation energy exchange can be expressed as $T(r,t) = T_{0} + \Delta T(r) \left[1 - \exp(-t/\tau_{b})\right]$, where $\tau _{b}$ is the decay constant of the birefringence signal, $T_{0}$ is the ambient temperature and $\Delta T(r)$ is the temperature increase at the end of the thermalization process. The dependence of the latter on $r$ reflects the gaussian profile of the centrifuge beam.

This work has been supported by the grants from CFI, BCKDF and NSERC. We are grateful to Ilya Sh. Averbukh for many stimulating discussions af the rotational effects on gas hydrodynamics. We thank Kirk W. Madison for bringing our attention to the schlieren imaging technique.


\end{document}